\documentclass{article}

\usepackage{graphicx}

\title{The Spread of Opinions and Proportional Voting}
\author{Gonzalo Travieso and Luciano da Fontoura Costa\\
Instituto de F\'{\i}sica de S\~ao Carlos, USP\\
Av. do Trabalhador S\~ao-carlense, 400\\
13566-590, S\~ao Carlos, SP,
Brazil\\
\{gonzalo,luciano\}@ifsc.usp.br}

\begin{document}

\maketitle

\begin{abstract}
  Election results are determined by numerous social factors that
  affect the formation of opinion of the voters, including the network
  of interactions between them and the dynamics of opinion influence.
  In this work we study the result of proportional elections using an
  opinion dynamics model similar to simple opinion spreading over a
  complex network. Erd\H{o}s-R\'enyi, Barab\'asi-Albert, regular
  lattices and randomly augmented lattices are considered as models of
  the underlying social networks. The model reproduces the power law
  behavior of number of candidates with a given number of votes found
  in real elections with the correct slope, a cutoff for larger number
  of votes and a plateau for small number of votes. It is found that
  the small world property of the underlying network is fundamental
  for the emergence of the power law regime.
\end{abstract}

\section{Introduction}
\label{sec:intro}

There has been a growing interest in the study of social phenomena
through the use of tools from statistical physics
\cite{Forsythe1996,Stanley1999,Stauffer2000,Stauffer2003}.  This
trend has been in part stimulated by developments in complex networks
\cite{Barabasi2002,Dorogovtsev2002,Newman2003,Boccaletti2006}, which
have uncovered properties of the structures underlying the
interactions between agents in many natural, technological, and social
systems. Social processes can be simulated through the use of complex
networks models over which a dynamical interaction between the agents
represented by the nodes is defined, yielding results that can be
compared with the macroscopic results found in real social networks.

Election of representatives are important social processes in
democracies, where a large number of people take part and that
represent the result of many social factors. It was found
\cite{Costa1999} that the number of candidates with a given number of
votes in the 1998 Brazilian elections follows a power law with slope
$-1$ for some orders of magnitude, or a generalized Zipf's law
\cite{Lyra2003}.

Elections depend on the process of opinion formation by the voters.
Each voter chooses one candidate based on its beliefs and through
interaction with other voters. Many works have been carried out on
opinion formation while considering several types of dynamics and
underlying network topologies.  Bernades \emph{et al.}\
\cite{Bernardes2002} and Gonz\'alez \emph{et al.}\ \cite{Gonzalez2004}
succeded in reproducing the general $-1$ slope of candidates with a
given number of votes in Brazilian election results by using the
Sznajd \cite{Sznajd2000} opinion formation model adapted to complex
networks.

In the Sznajd model, two neighbors that happen to have the same
opinion may convince their other neighbors. In this article, we adopt
a simpler model, where each single voter tries to convince its
neighbors, regardless of their previous opinion. The obtained results
exhibited a substantial agreement with real election results for some
network models.

The article is organized as follows. Firts we describe the network
(Sec.~\ref{sec:nets}) and opinion (Sec.~\ref{sec:opinion}) models used
in the simulations. Then, in Sec.~\ref{sec:results} we present and
discuss the simulation results and study the effect of the model
parameters. Finally, the conclusions are summarized in
Sec.~\ref{sec:conclusions}.

\section{Opinion and Network Models Used}

As done in other related works, we assume that the opinion formation
for the voting process occurs as interactions between agents connected
through a complex network. The result is thus determined by two
factors: (i)~the structure of the network that specify the possible
interactions between agents, and (ii)~the dynamics of opinion
formation between interacting agents. The following subsections
describe the models used in this work.

\subsection{Network Models}\label{sec:nets}

The voters and their social interactions are represented as a network,
so that the individuals are represented by nodes in the network and
every social interaction between pairs of voters is represented by a
link between the two corresponding nodes. The number of links attached
to a node is called the \emph{degree} of the node; the social distance
between to voters is given by the geodesic distance in the network,
defined as the minimum number of links that must be traversed in order
to reach one of the nodes starting from the other. Two important
network properties~\cite{Newman2003} are the degree distribution and
the average distance between pairs of nodes.

For the simulation of the opinion formation model we adopted the
Erd\H{o}s-R\'enyi and the Barab\'asi-Albert \cite{Barabasi2002} models
of complex networks. For comparison, simulations were also performed
in two-dimensional lattices and two-dimensional lattices with random
connections added between its nodes. The Erd\H{o}s-R\'{e}nyi networks
are characterized by a Poisson degree distribuion and the presence of
the ``small world'' property: the average distance between nodes grows
slowly with the number of nodes in the network. The Barab\'asi-Albert
model also has the small world property, but its degree distribuition
follows a power law, resembling in that sense many social
networks. The regular lattice was chosen as an example of a network
without the small world property, while the addition of random
connections enables a controled introduction of this property
(see~\cite{Watts1998}).

In the Barab\'asi-Albert model, the network starts with $m+1$
completely connected nodes and grows by the successive addition of
single nodes with $m$ connections established with the older nodes,
chosen according to the preferential attachment rule.  The growth
stops when the desired number of nodes $N$ is reached.

To generate the Erd\H{o}s-R\'enyi network, we start with $N$ isolated
nodes and insert $L$ links connecting pairs of nodes chosen with
uniform probability, avoiding self- and duplicate connections; for
comparison with the Barab\'asi-Albert model, we choose $L$ so that
$m=L/N$ is the same as the $m$ values used for the Barab\'asi-Albert
model.

For the two-dimensional lattices, the $N$ nodes are distributed in a
square and the connections are established between neighboring nodes
in the lattice.  Afterwards, additional connections can be
incorporated between uniformly random chosen pairs of nodes until a
desired number of average additional links per node is included. This
kind of randomly augmented regular network is similar to that used in
Newman and Watts small-world model \cite{Newman1999}.

\subsection{Opinion Model}\label{sec:opinion}

For a given network with $N$ voters (nodes), we start by distributing
the $C$ candidates among randomly chosen nodes (with uniform
probability), that is, each candidate is assigned to just one node in
the network (this reflects the fact that the candidates are also
voters). The remaining voters start as ``undecided'', meaning that
they have no favorite candidate yet.  The following process is
subsequently repeated a total of $S N$ times: choose at random a voter
$i$ that already has an associated candidate $c_i$; for \emph{all}
neighbors of voter $i$, if they have no associated candidate (i.e. are
as yet undecided), they are associated with candidate $c_i$, otherwise
they change to candidate $c_i$ with a given \emph{switching
probability} $p$. The constant $S$ introduced above is henceforth
called the \emph{number of steps} of the algorithm (average number of
interactions of each node). This opinion model is motivated by the
following assumptions: (i)~undecided voters are passive, in the sense
that they do not spread their lack of opinion to other voters;
(ii)~undecided voters are easily convinced by interaction with someone
that already has a formed opinion; (iii)~the flexibility to change
opinions due to an interaction, quantificed by the parameter $p$, is
the same for all voters. Despite the many limitations which can be
identified in these hypotheses, they seem to constitute a good first
approximation an can be easily generalized in future works.

This model is similar to a simple spreading to unoccupied sites, and
can be reduced to an asynchronous spreading if the switching
probability is zero. In spite of its simplicity, the model yields
interesting results, as discussed below.

\section{Results}\label{sec:results}

In the following, we present and discuss the histograms expressing the
number of candidates with a given number of nodes. The plots are in
logarithmic scale, and the bin size doubles from one point to the next
in order to provide uniformity. The number of candidates in a bin are
normalized by the bin size. All results correspond to mean values
obtained after $30$ different realizations of the model with the given
parameters.

As becomes clear from an analysis of the following graphs, larger
values of $N/C$ tend to lead to more interesting results, motivating
the adoption of large $N$ and small $C$. The use of too large values
of $N$ implies a high computational and memory cost; the use of too
small values of $C$ leads to poor statistics implied by the large
variations in the number of candidates inside the bins. The standard
values of $N=2\,000\,000$ and $C=1\,000$ adopted in the following
represent a good compromise considering our computational resources.

\begin{figure}
  \includegraphics[width=0.48\textwidth]{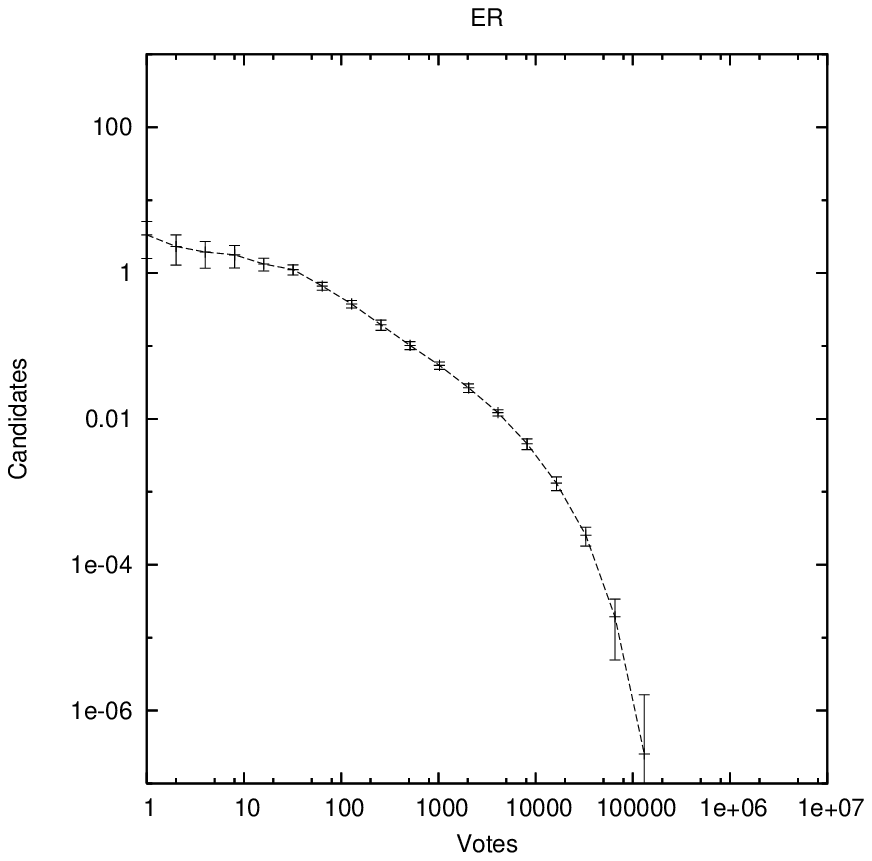}
  \includegraphics[width=0.48\textwidth]{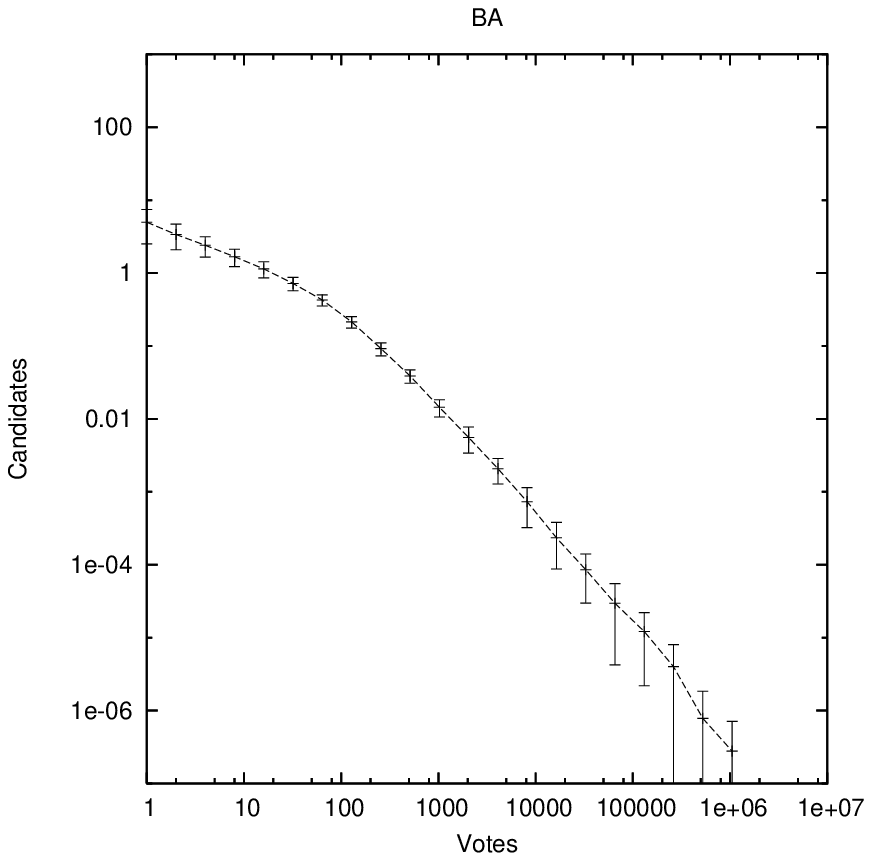}
  \caption{Distribution of candidates with a given number of votes
  after $30$ steps for networks with $2\,000\,000$ voters, $1\,000$
  candidates, $5$ links per node and a switching probability of
  $0.1$. On the lefthand side for Erd\H{o}s-R\'enyi and on the
  righthand side for Barab\'asi-Albert networks. Error bars show one
  standard deviation.}  \label{fig:errors}
\end{figure}

Figure~\ref{fig:errors} shows the results of the simulation for
Erd\H{o}s-R\'enyi and Barab\'asi-Albert networks after $30$ steps and
with a switching probability of $0.1$. The result for the
Erd\H{o}s-R\'enyi network is very similar to results of real elections
\cite{Costa1999}. There is a power-law regime for intermediate number
of votes, a plateau for small number of votes and a cutoff for large
number of votes; the power-law regime has an exponent of $-1$, which
is almost the same as that obtained for real
elections~\cite{Costa1999}. The large variability on the plateau
region is also consistent with the differences found at this part of
the curves when considering different elections outcomes (see for
example the data in \cite{Lyra2003}).

For the Barab\'asi-Albert model, although two power-law regimes with
different exponents can be identified, neither corresponds to the
experimental value of $-1$.

\begin{figure}
  \includegraphics[width=0.48\textwidth]{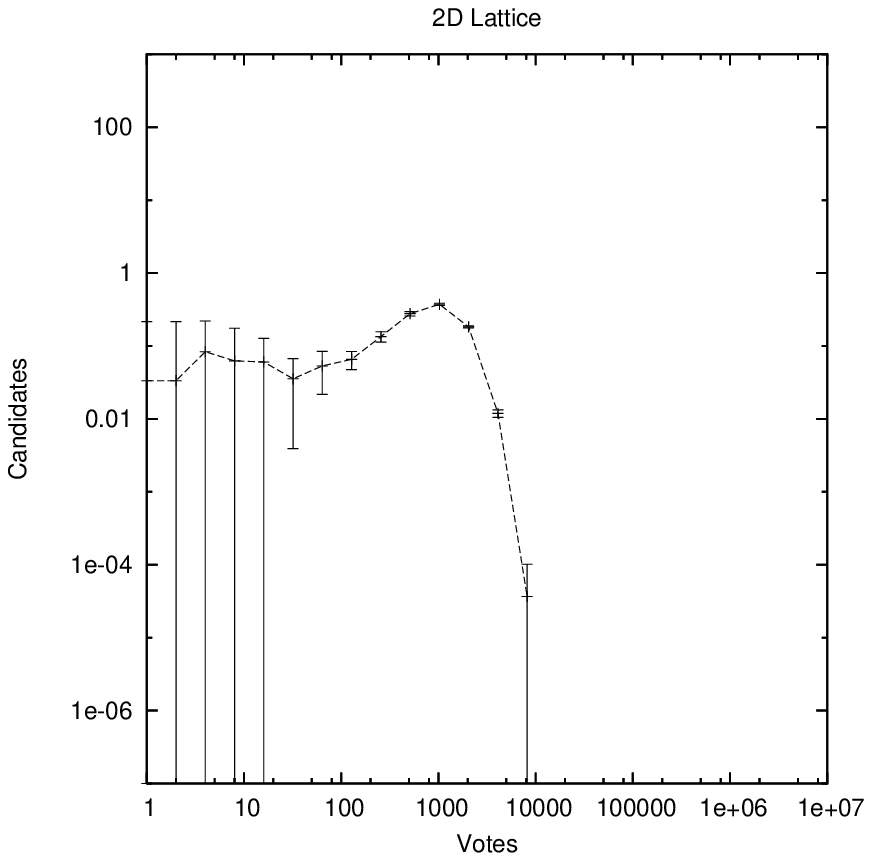}
  \includegraphics[width=0.48\textwidth]{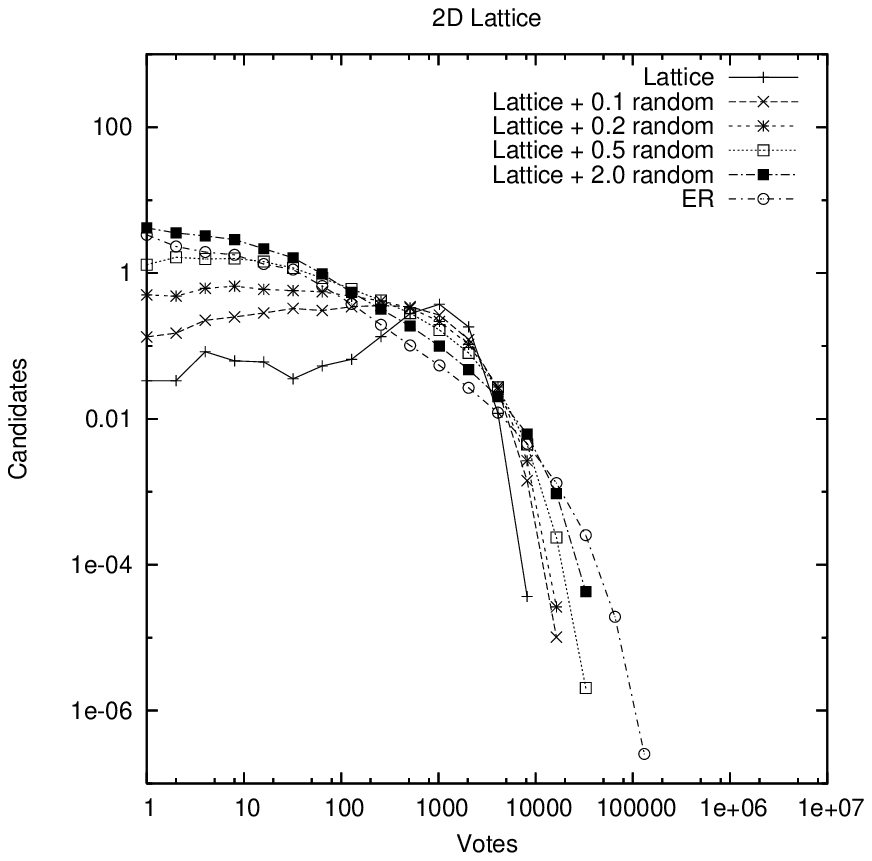}
  \caption{Distribution of candidates with a given number of votes
  after $30$ steps for two-dimensional lattices with $2\,000\,000$
  voters, $1\,000$ candidates, $5$ links per node and a switching
  probability of $0.1$. On the lefthand side for a pure lattice (error
  bars show one standard deviation) and on the righthand side for
  lattices with the addition of the given average number of shortcut
  links per node between randomly selected nodes. The result for the
  Erd\H{o}s-R\'enyi network is also shown for comparison.}
 \label{fig:lattice}
\end{figure}

The lefthand side of Figure~\ref{fig:lattice} shows the result for the
simulation on a two-dimensional lattice. There is no sign of a
power-law regime and a clear peak around $1\,000$ votes can be noted,
in disagreement with the scale-free nature of the experimental
results.  On the righthand side of the same figure, the effect of
adding random connections to the lattice can be easily visualized. It
is remarkable that the addition of just a small number of new links
(about half the number of nodes) is enough to get a result similar to
the one of the Erd\H{o}s-R\'enyi model. It is a known fact
\cite{Watts1998} that a small number of random links in a regular
network are enough to the emergence of the ``small world''
phenomenon. By enabling a candidate to reach the whole network of
voters in a small number of steps, this phenomenon increases the
chance of a candidate getting a very large number of votes, therefore
broadening the distribution.

\begin{figure}
  \includegraphics[width=0.48\textwidth]{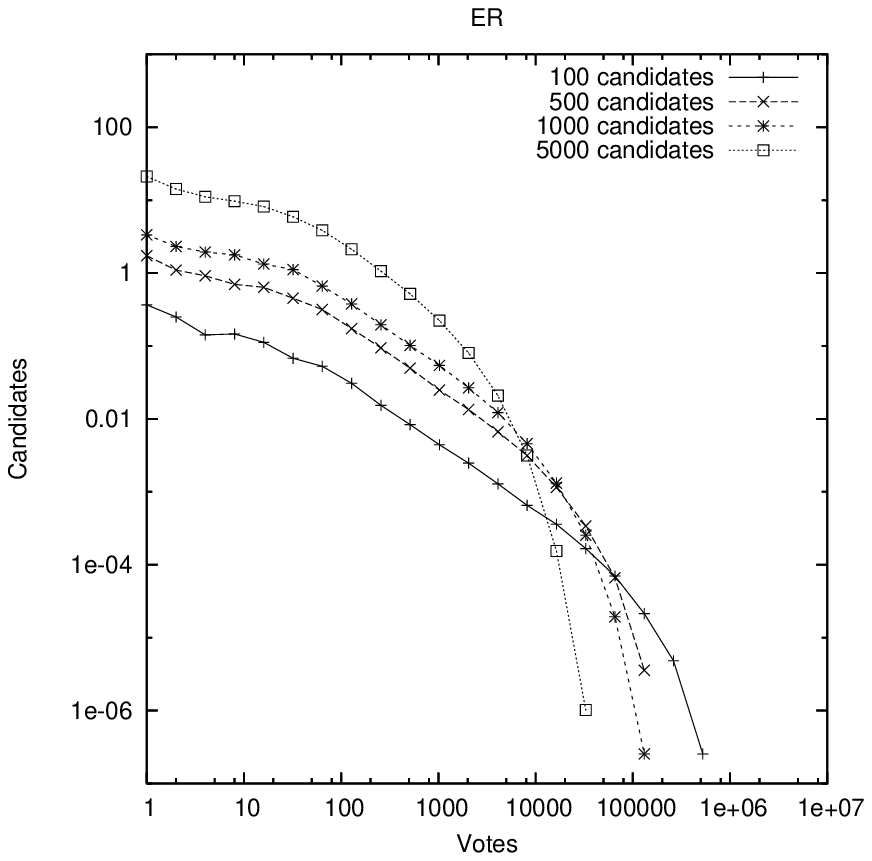}
  \includegraphics[width=0.48\textwidth]{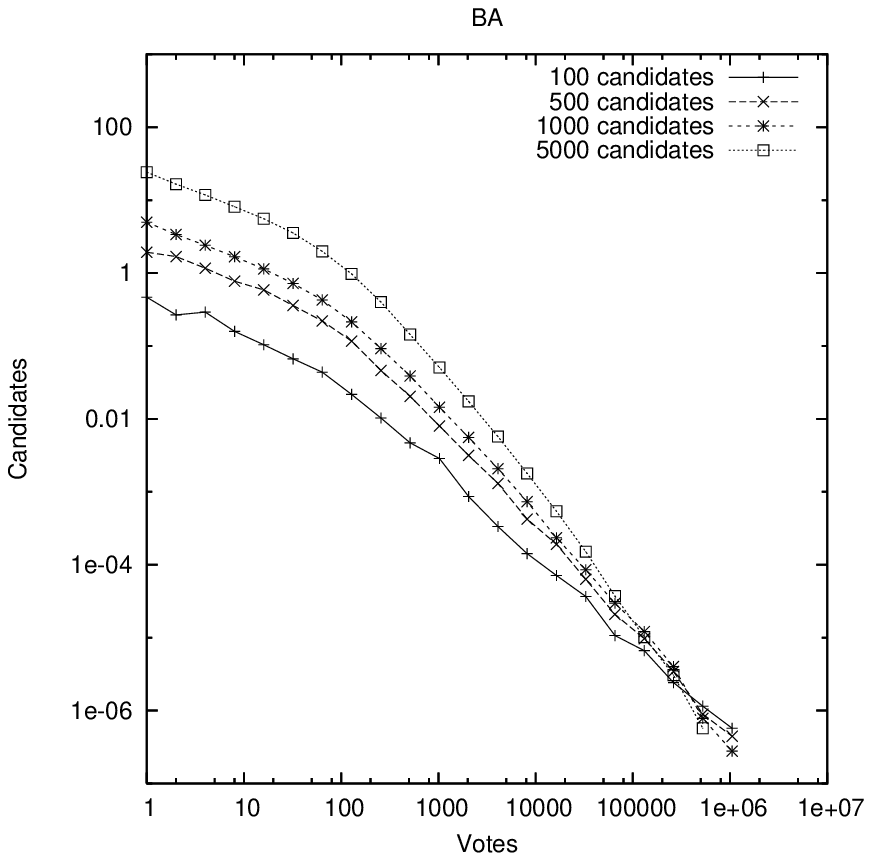}
  \caption{Effect of the number of candidates. Distributions after
    $30$ steps for networks with $2\,000\,000$ voters, $5$ links per
    node, a switching probability of $0.1$, and different number of
    candidates. On the lefthand side for Erd\H{o}s-R\'enyi and on the 
    righthand side for Barab\'asi-Albert networks.}
  \label{fig:cands}
\end{figure}

Now we turn our attention to the influence of the parameters of the
model. In Figure~\ref{fig:cands} the effect of changing the number of
candidates while keeping the other parameters fixed is shown. For the
Erd\H{o}s-R\'enyi model, the effect of increasing the number of
candidates translates itself as an upward shift of the curve while, at
the same time, the cutoff is shifted to the left. This is an expected
result: as the number of candidates grows with a fixed number of
voters, the candidates are initially distributed closer to one another
in the network, and have therefore fewer opportunities to spread
influence before hitting a voter already with an opinion; this leads
to a cutoff in smaller number of votes and in an increase in the
number of candidates with less votes than the cutoff. In the
Barab\'asi-Albert model, the behavior for small number of votes is
similar: the curve is shifted up; but for the power-law regime of
large number of votes, the curve decays more steeply as more
candidates are added.

\begin{figure}
  \includegraphics[width=0.48\textwidth]{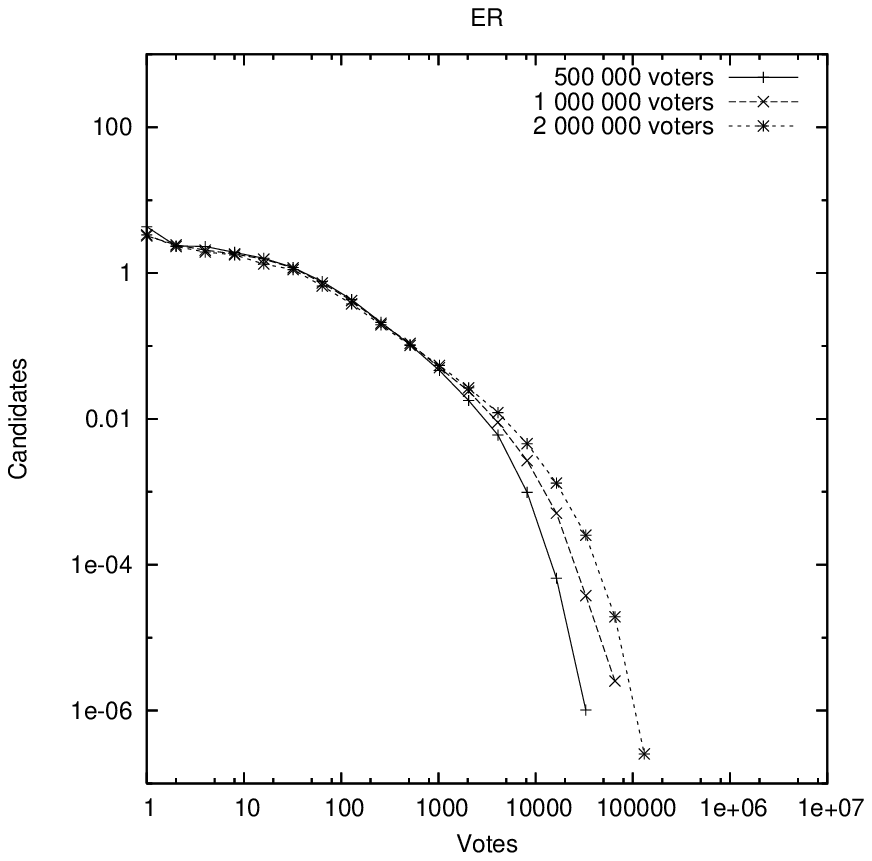}
  \includegraphics[width=0.48\textwidth]{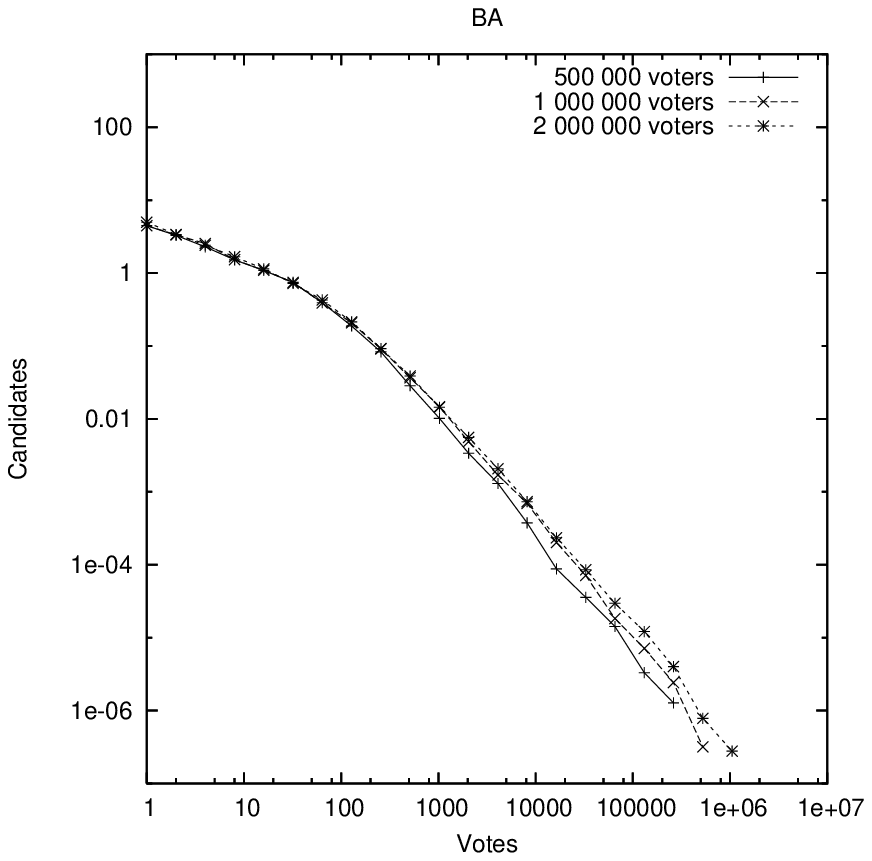}
  \caption{Effect of the number of voters. Distributions after $30$
    steps for networks with $1\,000$ candidates, $5$ links per node, a
    switching probability of $0.1$, and different number of voters. On
    the lefthand side for Erd\H{o}-R\'enyi and on the righthand side for
    Barab\'asi-Albert networks.}
  \label{fig:voters}
\end{figure}

Changing the number of voters has an impact limited almost exclusively
to the tail of the curves, as seen in Figure~\ref{fig:voters}. When
the number of voters is increased, in the Erd\H{o}s-R\'enyi model, the
cutoff is shifted to the left and the power-law regime is
correspondingly increased. In the Barab\'asi-Albert model, the maximum
number of votes is shifted and the inclination of the second power-law
regime is changed to acomodate this displacement. Comparing with
Figure~\ref{fig:cands}, we see that the tail of the curve for the
Barab\'asi-Albert model adapts its inclination according to the
relation between number of voters and candidates, i.e. a larger value
of $N/C$ implies a flatter tail.

\begin{figure}
  \includegraphics[width=0.48\textwidth]{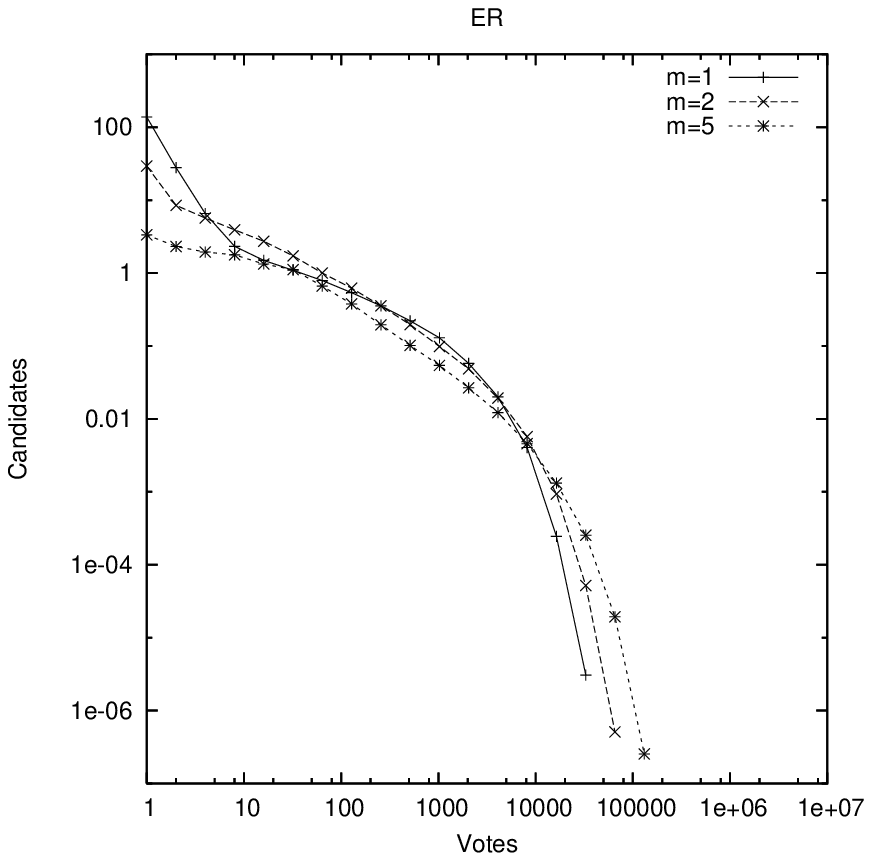}
  \includegraphics[width=0.48\textwidth]{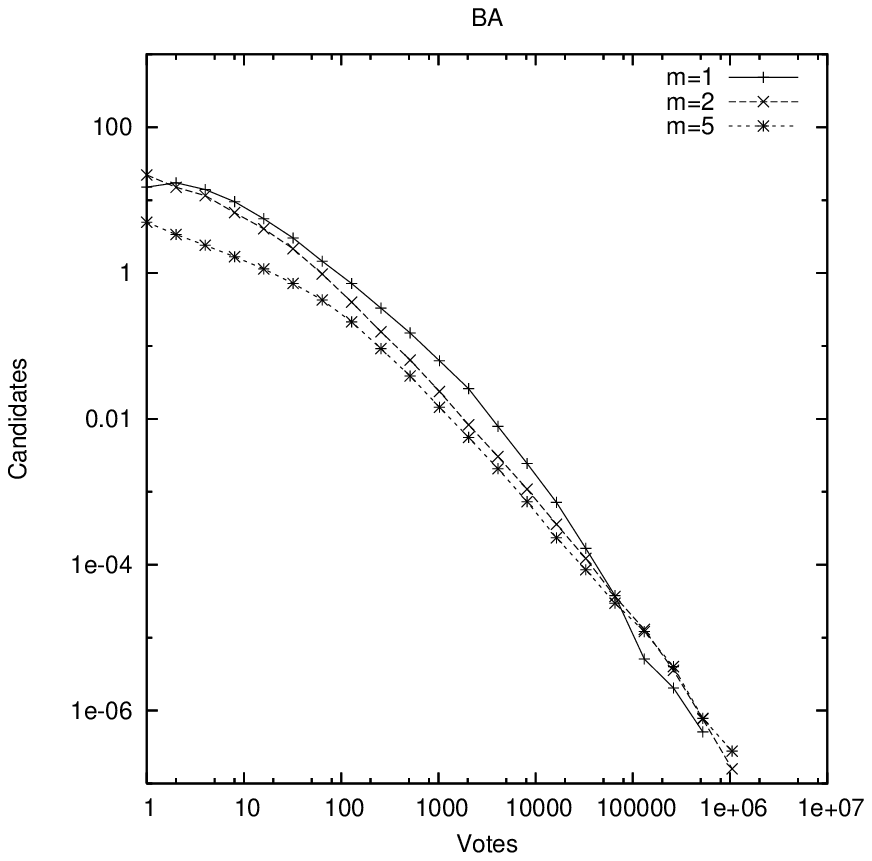}
  \caption{Effect of the number of links. Distributions after $30$
    steps for networks with $2\,000\,000$ voters, $1\,000$ candidates,
    a switching probability of $0.1$, and different number of links
    per node. On the lefthand side for Erd\H{o}s-R\'enyi and on the 
    righthand side for Barab\'asi-Albert networks.}
  \label{fig:links}
\end{figure}

From Figure~\ref{fig:links} we can see that the behavior that is being
discussed appears only if the network is sufficiently connected: for
$m=1$ there is no power-law regime for the Erd\H{o}s-R\'enyi model and
the behavior for the Barab\'asi-Albert model is complex, with three
different regions and a peak for small number of votes. Also for this
latter model, the inclination of the tail of the curve appears to be
slightly influenced by the average connectivity, with steeper tails
for smaller connectivities.

\begin{figure}
  \includegraphics[width=0.48\textwidth]{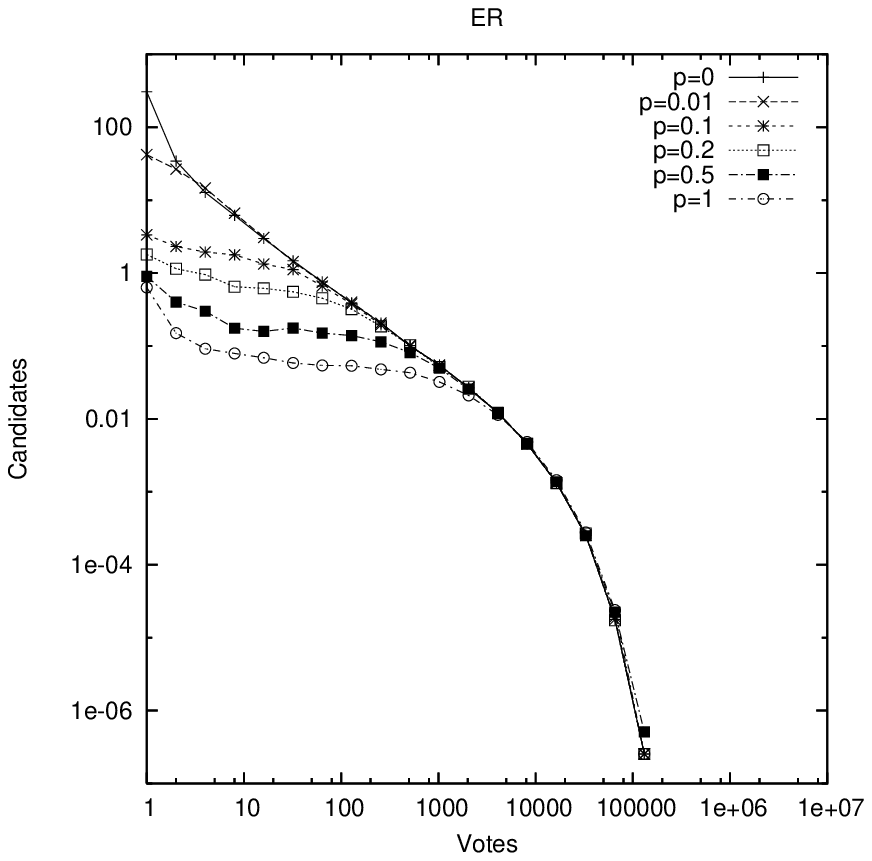}
  \includegraphics[width=0.48\textwidth]{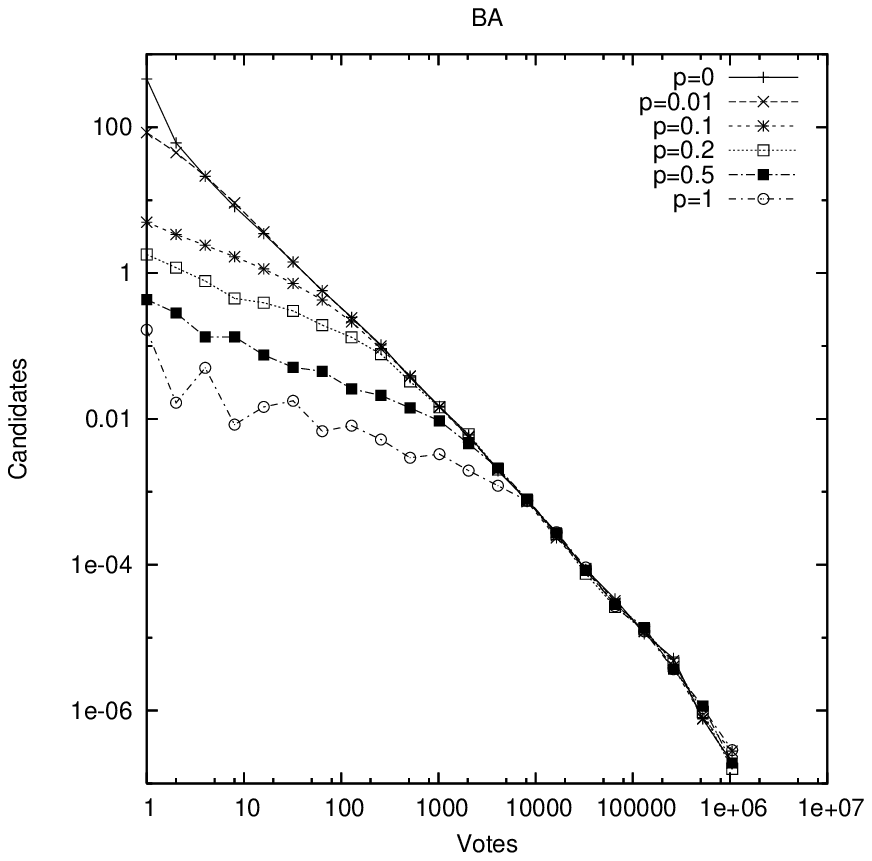}
  \caption{Effect of the swtiching probability. Distributions after
    $30$ steps for networks with $2\,000\,000$ voters, $1\,000$
    candidates, $5$ links per node, and different values for the
    switching probability. On the lefthand side for Erd\H{o}s-R\'enyi 
    and on the righthand side for Barab\'asi-Albert networks.}
  \label{fig:change}
\end{figure}

The switching probability has effect only on the first part of the
curve, as can be seen from Figure~\ref{fig:change}. In both models,
this part of the curve is shifted down as the probability increases
and its range is extended until it touchs the original (for zero
probability) curve. Note that the inclination of the Barab\'asi-Albert
curve corresponding to small number of votes is maintained for the
different values of switching probability (but is different for zero
probability).

\begin{figure}
  \includegraphics[width=0.48\textwidth]{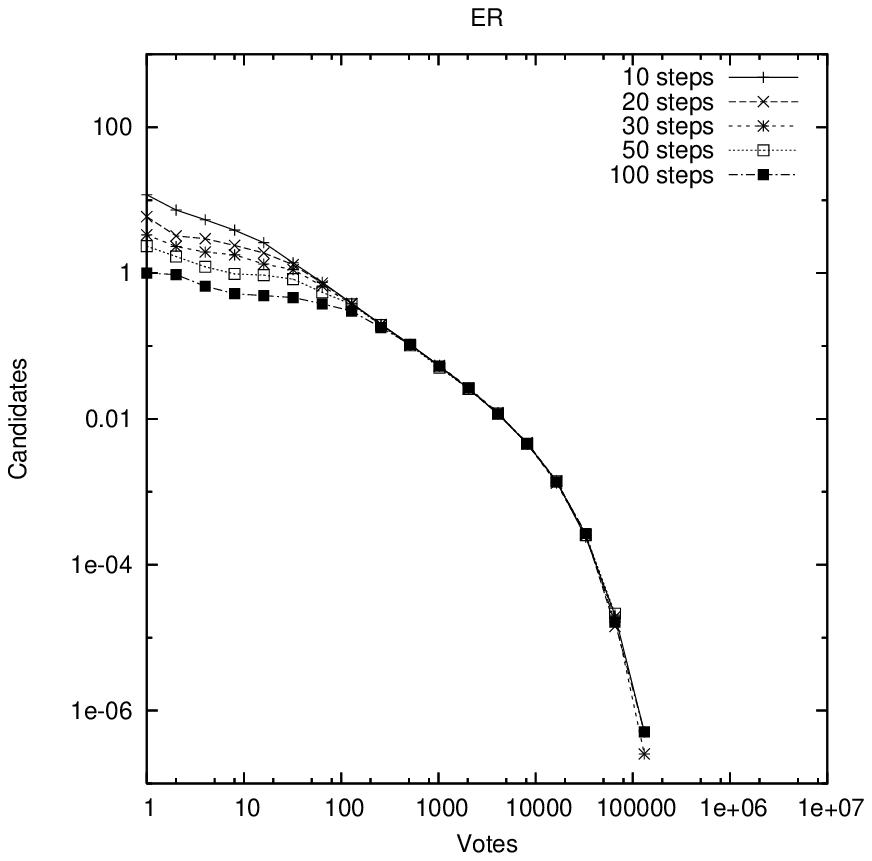}
  \includegraphics[width=0.48\textwidth]{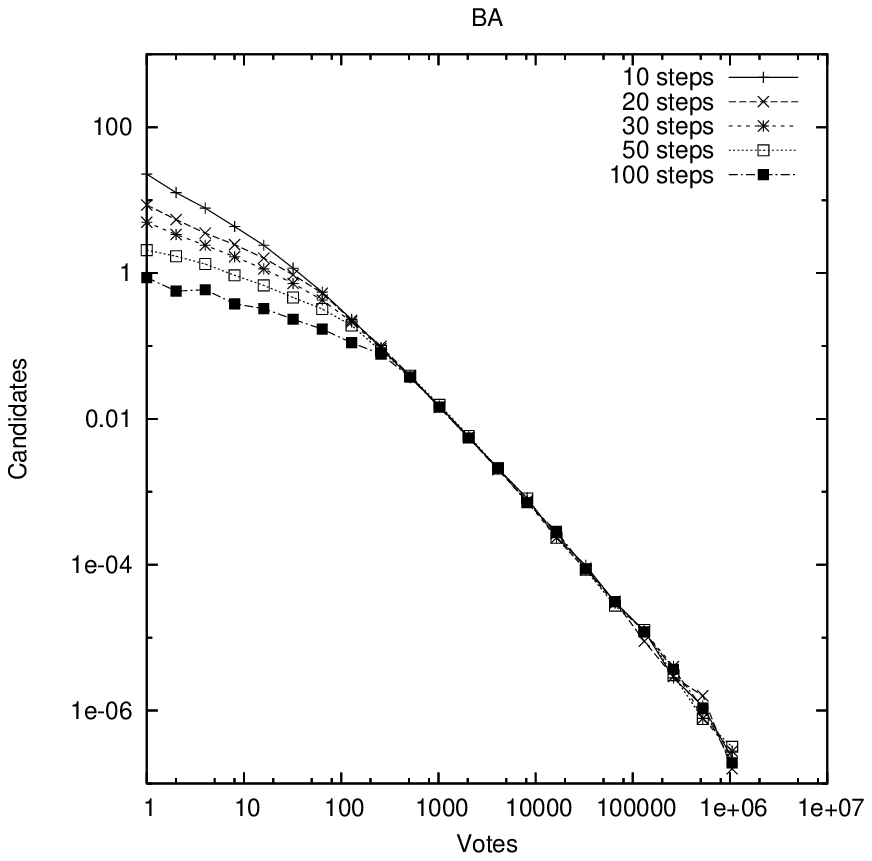}
  \caption{Effect of the number of steps. Distributions for networks
    with $2\,000\,000$ voters, $1\,000$ candidates, $5$ links per
    node, a switching probability of $0.1$, and different total number
    of steps. On the lefthand side for Erd\H{o}s-R\'enyi and on the 
    righthand side for Barab\'asi-Albert networks.}
  \label{fig:steps}
\end{figure}

A similar effect has been obtained while changing the number of steps
(Figure~\ref{fig:steps}). As the number of steps is increased, the
curve remains unchanged for large number of votes, but is donwshifted
for small number of votes. The similarity between an increase in the
number of steps and an increase in switching probability is easyly
explained: after all voters have a candidate, changes occur only by
switching candidates.  In other words, increasing the number of steps
has as an effect increase in the number of times a switching is tried,
resulting in a similar effect as increasing the switching probability.

\section{Conclusions}
\label{sec:conclusions}

We suggested and studied a simple voting model based on the spreading
of opinions through the links of a network. The results of the
simulation of the model show a remarkable qualitative agreement with
experimental results for proportinal voting in Brazilian and Indian
elections \cite{Costa1999} when the network model used is of
Erd\H{o}s-R\'enyi type or a lattice with sufficient random shortcuts
added. In these networks, the model results in a power-law
distribution with an exponent of $-1$, but with a shortcut for large
number of votes and a plateau for small number of votes, as observed
in real elections. The ``small world'' effect appears to be of central
importance in this result, as the result for a lattice without
shortcuts is very different, without any power-law regime.

Interestingly, the Barab\'asi-Albert network model gives results that
are not consistent with real elections: there are two power-law
regimes without a shortcut and the second (and dominant) power-law
regime is not universal, depending on the number of links per node in
the network and the relation between number of voters and number of
candidates. Also, the first power-law regime is not characterized by
the experimental value of $-1$. This is somewhat puzzling, as many
social networks have power-law degree distribuitions \cite{Newman2003}
and are in this respect better related to the Barab\'asi-Albert model
than to the other two models investigated. We suspect the explanation
to this is related to the importance of clustering and communities in
social networks, neither of which represented in the Barab\'asi-Albert
model, although they are not present also in the Erd\H{o}s-R\'enyi
networks. This in an issue deserving further investigation.

\vspace{0.5cm}
\noindent {\bf Acknowledgements:} L. da F. Costa is grateful to CNPq
(308231/03-1) for financial sponsorship.

\bibliographystyle{unsrt}
\bibliography{spreading}

\end{document}